**Reply to "Comment on 'Apical charge flux-modulated in-plane transport properties of cuprate superconductors'"**

The preceding Comment [1] mainly claims that the correlation of $T_{c,max}$ versus apical force described in Figs. 1(a), 1(b) of Ref. [2] can be included in a correlation in Ref. [3]. Their statements misunderstood our entire argument, which is thus misleading. The apical force is calculated by perturbation of apical ions, which was designed by us as a generator of directional charge fluxes and a measure of restoring force constants for lattice oscillations. Our correlation of $T_{c,max}$ versus apical force thus leads to the critical discussion about charge fluxes coupled with oxygen oscillations (Fig. 2 [2]), and the effect of cooperative charge fluxes of neighboring sites on the in-plane hole hopping behavior (Fig. 3 [2]). Ours is thus a dynamic picture that explicitly couples small charge fluxes, ion oscillations and the hole hopping on an ultrafast timescale, where the force correlation is just a starting point. On the contrary, their emphasis of axial orbital energy $\varepsilon_s$ still reflects a static picture with fixed ions and without charge fluxes. Although there certainly could be correlations between apical force, apical bonding and $\varepsilon_s$, which is not surprising as they all were related to electronic structures along $dz^2$ direction extracted from DFT simulations of cuprates, we want to emphasize here that the two pictures are completely different in terms of dynamic versus static. For example, we prefer stronger dynamic apical flux modulation for higher $T_{c,max}$, whereas they prefer weaker static apical modulation in $\varepsilon_s$.

Furthermore, they clarify that we should not give people the impression that their $T_{c,max}$ correlation is *simply* with the apical oxygen height $d_A$ discussed in Fig. 4 in Ref. [3], because they also mentioned some other descriptor. This is a misinterpretation of our statement, because, first, we argued in Fig. 1d that if considering all cuprate families, there is *not* a clear correlation of $T_{c,max}$ with the apical oxygen height, and, second, in the introduction we mentioned both their $d_A$ and $t'/t$ correlations. However, I notice that at a different place of their Comment, they also wrote "$T_{c,max}$ generally increases with $d_A$ is not true", which is a statement that cannot be found in their original Ref. [3]. Nevertheless, we find that their picture still prefers larger apical oxygen height for higher $T_{c,max}$ (Fig. 4 [3]), which they claim also correlates with weaker static apical modulation $\varepsilon_s$, with $t^2_{AA,Cu\ 4s} \sim 0$ for the extreme case of Hg and Tl families in their Comment. Such extreme case in our picture, however, is with the strongest dynamic $t^{flux}_{AA,Cu}$. A further development of the collective behavior of charge fluxes in our recent work [4] leads to a new picture of pseudogap phase, where certain conservation property of dynamic charge fluxes originates intertwined pseudogap and superconducting phases. We thus believe our charge flux process is not a "high-order" effect as they thought, but rather a critical component for cuprate superconductivity that their picture lacks.

They also wrote that "most importantly, whereas the correlation [3] extends in a straightforward way to the numerous higher-$T_c$ bi- and tri-layer cuprates, the correlation [2] does not". We have to say that for such a complicated topic of cuprate superconductivity, the "straightforward way" is not necessarily superior than a sophisticated way. We discussed the layer dependence in Fig. 2d [2], using again the charge flux picture, where increasing number of $CuO_2$ layers helps effectively screen the perturbation to the in-plane hole hopping process by suppressing the competition in flow directions of fluxes driven by the oscillation from the two opposite apical sides. Such perturbation is thus the largest in the single layer case, which decreases with increasing $CuO_2$ layers. Based on our picture, we believe that the correlation presented in our Fig. 1a,b [2] should not include the layer dependence, which we chose to present separately in Fig. 2d [2]. Therefore, it is not appropriate to state that our work "fails to produce figures" that they prefer, to put both materials and layer dependences together in one plot. The different ways of presenting the layer dependence between the two works also reflect the fundamental difference in the two pictures. They argue the multilayer as a means of lowering the apical static modulation, which shares the same origin as their previous argument of weaker $\varepsilon_s$ for higher $T_{c,max}$, while we ascribe the multilayer as a mean of screening the competition in dynamic apical modulations, which is a mechanism works universally for either strong or weak apical fluxes.

Xin Li[1]*, Sooran Kim[1,2], Xi Chen[1], William Fitzhugh[1]
[1]School of Engineering and Applied Sciences, Harvard University, Cambridge, MA 02138, USA
*lixin@seas.harvard.edu
[2]Department of Physics Education, Kyungpook National University, Daegu, 41566, South Korea